\documentclass[aps,prd,twocolumn,preprintnumbers,superscriptaddress,floatfix]{revtex4}

\usepackage{multirow, graphicx,amssymb,url,mathrsfs,amsmath}
\usepackage{eucal,wrapfig,boxedminipage,setspace,subfigure}
\usepackage{amsxtra,amstext,latexsym,dsfont}

      \def\L  {\Lambda}

\def\IR{{\hbox{{\rm I}\kern-.2em\hbox{\rm R}}}}
\def\IB{{\hbox{{\rm I}\kern-.2em\hbox{\rm B}}}}
\def\IN{{\hbox{{\rm I}\kern-.2em\hbox{\rm N}}}}
\def\IC{\,\,{\hbox{{\rm I}\kern-.59em\hbox{\bf C}}}}
\def\IZ{{\hbox{{\rm Z}\kern-.4em\hbox{\rm Z}}}}
\def\IP{{\hbox{{\rm I}\kern-.2em\hbox{\rm P}}}}
\def\IH{{\hbox{{\rm I}\kern-.4em\hbox{\rm H}}}}
\def\ID{{\hbox{{\rm I}\kern-.2em\hbox{\rm D}}}}

\newcommand{\beq}{\begin{equation}}
\newcommand{\eeq}{\end{equation}}
\newcommand{\bea}{\begin{eqnarray}}
\newcommand{\eea}{\end{eqnarray}}

\begin{document}

\voffset 1cm

\newcommand\sect[1]{\emph{#1}---}

\title{Hyper-Scaling Relations in the Conformal Window from Dynamic AdS/QCD}

\author{Nick Evans \& Marc Scott}
\affiliation{STAG Research Centre and  Physics \& Astronomy, University of
Southampton, Southampton, SO17 1BJ, UK.\\ evans@soton.ac.uk, ms17g08@soton.ac.uk}

\begin{abstract}
 Dynamic AdS/QCD is a holographic model of strongly coupled gauge theories with the dynamics included through the running anomalous dimension of the quark bilinear, $\gamma$. We apply it to describe the physics of massive quarks in the conformal window of SU($N_c$) gauge theories with $N_f$ fundamental flavours, assuming the perturbative two loop running for $\gamma$. We show that to find regular, holographic, renormalization group flows in the infra-red the decoupling of the quark flavours at the scale of the mass is important and enact it through suitable boundary conditions when the flavours become on shell. We can then compute the quark condensate and the mesonic spectrum ($M_\rho, M_\pi, M_\sigma$) and decay constants. We compute their scaling dependence on the quark mass for a number of examples. The model matches perturbative expectations for large quark mass and na\"ive dimensional analysis (including the anomalous dimensions) for small quark mass. The model allows study of the intermediate regime where there is an additional scale from the running of the coupling and we present results for the deviation of scalings from assuming only the single scale of the mass.   
\noindent

\end{abstract}

\maketitle

\newpage
\section{Introduction}

Dynamic AdS/QCD \cite{Alho:2013dka} is a holographic model \cite{Maldacena:1997re} of strongly coupled gauge theories with quark matter (other holographic models of similar physics can be found in \cite{others}). Its elements are chosen to retain much of the simplicity of the original AdS/QCD models \cite{Erlich:2005qh} but with sufficient input from top-down holographic models of brane probes 
\cite{Karch:2002sh,Babington:2003vm,Alvares:2012kr} to enable it to describe the dynamics of chiral condensation. The gauge dynamics itself is input through an ansatz for the running of the anomalous dimension, $\gamma$, of the quark mass or $\bar{q}q$ operator.  The quark mass is the only remaining free parameter as in QCD. The model then determines the chiral condensate which provides an IR cut off in the geometry. The $\sigma$ mode mass, $f_\sigma$, the $\rho$ mass, $f_\rho$, and the $\pi$ mass are then predictions.  The main power of the model is to study the dependence of the spectrum on the ansatz for the running of $\gamma$. In \cite{Alho:2013dka} the model was used to study the spectrum of walking gauge theories in the massless limit and it displayed the expected behaviours very simply: the transition is of holographic BKT type (showing Miransky exponential scaling)\cite{Miransky:1996pd}, the quark condensate is enhanced relative to $f_\pi$ \cite{Holdom:1981rm} and the $\sigma$ meson mass is suppressed \cite{Haba:2010hu} relative to the rest of the spectrum in the walking regime.

Here we wish to extend the model to describe the conformal window \cite{Caswell:1974gg, Ryttov:2007cx} of SU($N_c$) gauge theories with $N_f$ fundamental flavours. For a theory with quarks in the fundamental representation, asymptotic freedom sets in when 
$N_f<11N_c/2$. Immediately below that point, at least at large $N_c$, the two loop beta function enforces a perturbative infra-red (IR) fixed point 
\cite{Caswell:1974gg}. The fixed point behaviour is expected to persist into the non perturbative regime as $N_f$ is further reduced. This behaviour is seen in the two loop perturbative computation of the running of the coupling of $\alpha$ and hence $\gamma$ and we will use that ansatz here to model these theories. Of course as the fixed point leaves the perturbative regime this becomes just a sensible ``parametrization'' of the non-perturbative physics. 

At some critical value of the number of flavours, $N_f^c$, the coupling is expected to be strong enough to trigger chiral symmetry breaking by the formation of a quark anti-quark condensate (so called walking theories live just on the symmetry breaking side of that transition). Holographic models describe the quark condensate by a scalar in AdS whose mass is related to the mass dimension, $\Delta$, of the field theory operator  ($m^2=\Delta(\Delta-4)$)\cite{Maldacena:1997re}. As $\Delta$ falls through 2 (or $\gamma\geq1$), a clear instability sets in as the mass violates the Breitenlohner-Freedman (BF) bound in AdS$_5$\cite{Breitenlohner:1982jf}. Remarkably, the $\gamma=1$ criterion precisely matches that deduced from gap equation analysis of the same problem\cite{Appelquist:1988yc}. Using the two loop running for $\gamma$ the BF bound violation first occurs at $N_f \simeq 4 N_c$. 

In this paper we will mainly concentrate on the conformal window at values of $N_f$ above $N_f^c$. As an example we set $N_c=3$ ($N_f^c\simeq 11.9$) and look at a few descrete values of $N_f$ ($12$, $13$ and $15$) which span the conformal window regime. These examples suffice to explore the qualitative behaviour of observables on the running to different IR fixed point values of $\gamma$ and are easily extendable to different $N_c$. In the massless quark limit these theories flow to a non-trivial and strongly coupled IR conformal theory. The existence of such theories is of great theoretical interest and a sizeable lattice community \cite{lattice} is seeking evidence for them in numerical simulations. On a lattice the massless limit can only be obtained as a fine tuned point in parameter space. Simulations are therefore performed with finite mass and signals of the presence of, and approach to, the conformal phase are sought. For this reason a simple model such as ours, that makes predictions for this limit, should be helpful in identifying expected behaviours in physical, measurable quantities as one approaches the fixed point. 
We will therefore concentrate on studying the dependence of the quark condensate, meson masses and decay constants on the quark mass. 

We had previously only briefly considered the massive model. Here we show that the IR of the most na\"ive model does not allow regular flows for the scalar field describing the quark mass and condensate. The reason for this is that the model does not encode the decoupling of the quark flavours in the background geometry below their mass scale. We provide a simple fix for this issue - imposing an IR boundary condition at the scale where the mass-shell condition is satisfied.  It is then a straightforward matter to compute in the model. We first calculate the quark condensate as a function of quark mass, $m$. There is a leading divergence in the UV of the form $m \Lambda_{UV}^2$ as one would expect on dimensional grounds. In the IR conformal regime with non-zero  (and approximately constant) $\gamma$ this relation takes the form $m \Lambda_{UV}^{2-2\gamma}$. This matches na\"ive dimensional analysis - if the mass has dimension $1+\gamma$ and the condensate dimension $3- \gamma$ then we expect this dependence on the UV scale. There is then a sub-leading term in the condensate which grows as $m^3$ in the UV but as 
 $m^{3- \gamma \over 1 + \gamma}$ in the IR fixed point regime. This is again consistent with dimensional analysis in the IR - these are the hyperscalings relations found in \cite{DelDebbio:2010ze}. One of the powers of holographic descriptions is that the model reproduces this scaling.

Changing the precise IR boundary condition leaves these power relations invariant but changes the constant of proportionality between $\langle \bar{q} q \rangle$ and  $m^{3- \gamma \over 1 + \gamma}$. Once this constant is chosen the model allows one to follow the renormalization group flow of the mass and the condensate. Numerical work lets us look at intermediate regimes where the quark mass is of order the scale $\Lambda_1$ (roughly the scale generated by the 1-loop beta function)  where the coupling transitions from the perturbative regime to the non-perturbative fixed point.  To analyze the impact of $\Lambda_1$ we fit to a simple scaling relation of the form $m^b$. The exponent $b$ can then be translated, using the hyperscaling relation, to a predicted value for $\gamma$ which we compare to the functional form of $\gamma$ that we have input into the model. In the running regime around $\Lambda_1$ we find significant deviations from the input $\gamma$ showing that the one-scale $m^b$ functional form breaks down in the running regime. In this regime there is of course the second scale $\Lambda_1$ so this is as expected - our method allows us to quantify the deviations in this regime though. 

Most importantly for comparison to lattice simulations are computations of physical observables. We compute the meson spectrum including $M_\rho, M_\pi$ and $M_\sigma$ and their decay constants and display their scalings with $m$ and against each other.  When we compute these dimension 1 quantities  we expect hyper-scaling behaviour for dimension one objects of the form $m^{1/1+ \gamma}$. We can again extract $\gamma$ from each variable and display variations from the input $\gamma$ function in the different regimes. The hyper-scaling relations are matched in the deep UV and IR fixed point regimes but there are significant deviations in the running regime where $\Lambda_1$ again enters the physics. These are the main results of our analysis. 

\bigskip
\section{Dynamic AdS/QCD}

Dynamic AdS/QCD was introduced in detail in \cite{Alho:2013dka}. The model maps onto the action of a probe D7 brane in an AdS geometry expanded to quadratic order \cite{Alvares:2012kr}. The anomalous dimension of the quark mass/condensate is encoded through a mass  term that depends on the radial AdS coordinate $\rho$.

The five dimensional action of our effective holographic theory is
\bea
S & = & \int d^4x~ d \rho\, {\rm{Tr}}\, \rho^3 
\left[  {1 \over \rho^2 + |X|^2} |D X|^2 \right. \nonumber \\ 
&& \left.+  {\Delta m^2 \over \rho^2} |X|^2   + {1 \over 2} F_V^2  \right], 
\label{daq}
\eea
The  field $X$ describes
the quark condensate degree of freedom. Fluctuations in $|X|$ around its vacuum configurations will describe the scalar meson. The $\pi$ fields are the phase of $X$,
\begin{equation} X = L(\rho)  ~ e^{2 i \pi^a T^a} .
\end{equation}
$F_V$ are vector fields that will describe the vector ($V$)  mesons. It is possible to  include additional mesonic states through extra holographic fields that describe further QCD operators. For example, in \cite{Alho:2013dka} we included the a-mesons through an axial gauge field. The simpler model here though contains sufficient physical observables to display the behaviours we are interested in.  

We work with the five dimensional metric 
\begin{equation} 
ds^2 =  { d \rho^2 \over (\rho^2 + |X|^2)} +  (\rho^2 + |X|^2) dx^2, 
\end{equation}
which will be used for contractions of the space-time indices.
$\rho$ is the holographic coordinate ($\rho=0$ is the IR, $\rho \rightarrow \infty$ the UV)
and $|X|=L$ enters into the effective radial coordinate in the space, i.e. there is an effective $r^2 = \rho^2 + |X|^2$. This is how the quark condensate generates a soft IR wall for the linearized fluctuations that describe the mesonic states: when $L$ is nonzero the theory will exclude the deep IR at $r=0$.


The normalizations of $X$ and $F_V$ are determined by matching to the gauge theory in the UV. External currents are associated with the non-normalizable modes of the fields in AdS. In the UV we expect 
$|X| \sim 0$ and we can solve
the equations of motion for the scalar, $L= K_S(\rho) e^{-i q.x}$ and vector $V^\mu= \epsilon^\mu K_V(\rho) e^{-i q.x}$field. Each satisfies the same equation
\begin{equation}  \label{thing}
\partial_\rho [ \rho^2 \partial_\rho K] - {q^2 \over \rho} K= 0\,. \end{equation}
The UV solution  is
\begin{equation} \label{Ks}
K_i = N_i \left( 1 + {q^2 \over 4 \rho^2} \ln (q^2/ \rho^2) \right),\quad (i=S,V),
\end{equation}
where $N_i$ are normalization constants that are not fixed by the linearized equation of motion.
Substituting these solutions back into the action gives the scalar correlator $\Pi_{SS}$ and the vector correlator $\Pi_{VV}$. Performing the usual matching to the UV gauge theory requires us to set
\begin{equation} N_S^2 = N_V^2 = {N_c N_f \over 24 \pi^2 }.
\end{equation}

The vacuum structure of the theory can be determined by setting all fields except $|X|=L$ to zero. We assume that $L$ will have no dependence on the $x$ coordinates. The action for $L$  is given by
\begin{equation} \label{act} S  =  \int d^4x~ d \rho ~  \rho^3 \left[   (\partial_\rho  L)^2 +  \Delta m^2 {L^2  \over \rho^2 }   \right].
\end{equation}
If $\Delta m^2 =0$ then the scalar, $L$, describes a dimension 3 operator and dimension 1 source as is required for it to represent $\bar{q} q$ and the quark mass $m$. That is, in the UV the solution for the $L$ equation of motion is $L = m + \bar{q}q/\rho^2$. A non-zero $\Delta m^2$ allows us to introduce an anomalous dimension for this operator. If the mass squared of the scalar violates the BF bound of -4 ($\Delta m^2=-1$, $\gamma=1$) then  the scalar field $L$ becomes unstable and the theory enters a chiral symmetry breaking phase.

We will fix the form of $\Delta m^2$ using the two loop running of the gauge coupling in QCD which is given by
\begin{equation} 
\mu { d \alpha \over d \mu} = - b_0 \alpha^2 - b_1 \alpha^3,
\end{equation}
where
\begin{equation} b_0 = {1 \over 6 \pi} (11 N_c - 2N_f), \end{equation}
and
\begin{equation} b_1 = {1 \over 24 \pi^2} \left(34 N_c^2 - 10 N_c N_f - 3 {N_c^2 -1 \over N_c} N_f \right) .\end{equation}
Asymptotic freedom is present provided $N_f < 11N_c/2$. There is an IR fixed point with value
\begin{equation} \alpha_* = -b_0/b_1\,, \end{equation}
which rises to infinity at $N_f \sim 2.6 N_c$. 

The one loop result for the anomalous dimension of the quark mass is
\begin{equation} \gamma_1 = {3 C_2 \over 2\pi}\alpha, \hspace{1cm} C_2= {(N_c^2-1) \over 2 N_c} \,.  \end{equation}
So, using the fixed point value $\alpha_*$, the condition $\gamma=1$ occurs at $N_f^c \sim 4N_c$ (precisely $N_f^c = N_c \left( {100 N_c^2 - 66 \over 25 N_c^2 - 15}\right)$).

We will identify the RG scale $\mu$ with the AdS radial parameter $r = \sqrt{\rho^2+L^2}$ in our model. Note it is important that $L$ enters here. If it did not and the scalar mass was only a function of $\rho$ then, were the mass to violate the BF bound at some $\rho$, it would leave the theory unstable however large $L$ grew. Including $L$ means that the creation of a non-zero but finite $L$ can remove the BF bound violation leading to a stable solution. 

Working perturbatively from the AdS result $m^2 = \Delta(\Delta-4)$ we have
\begin{equation} \label{dmsq3} \Delta m^2 = - 2 \gamma_1 = -{3 (N_c^2-1) \over 2 N_c \pi} \alpha\, .\end{equation}
This will then fix the $r$ dependence of the scalar mass through $\Delta m^2$ as a function of $N_c$ and 
$N_f$. 

It's important to stress that using the perturbative result outside the perturbative regime is in no sense rigorous but simply a phenomenological parametrization of the running as a function of $\mu, N_c,N_f$ that shows fixed point behaviour. Similarly the relation between $\Delta m^2$ and $\gamma_1$ is a guess outside of the pertabation regime. Note that the holographic fixed point value for the anomalous dimension is given by solving $\Delta(\Delta-4)= \Delta m^2$ and the resultant $\gamma$ will not be the same as the fixed point in $\gamma_1$ away from the perturbative regime. Below we will display $\gamma(\mu)$ extracted from the holographic $\Delta$ and describe it as the ``input anomalous dimension"' - the discrepancy from $\gamma_1$ is not large even in the deep IR in the conformal window regime of $N_f$. 

\bigskip
\section{Scaling Behaviour of the Quark Condensate}

Firstly, we will study the vacuum structure of SU($3$) gauge theory with $N_f$ fundamental quarks in the conformal window range $12\leq N_f\leq 15$. These theories are conformal when the quarks are massless so  we will study the theories with a quark mass which breaks conformality. We will show that the model correctly encodes the running dimensions of the quark mass and condensate.

The Euler-Lagrange equation for the determination of $L$, in the case of a constant $\Delta m^2$, is 
\begin{equation} \label{embedeqn}
\partial_\rho[ \rho^3 \partial_\rho L]  - \rho \Delta m^2 L  = 0\,. \end{equation}
If $\Delta m^2$ depends on $L$ then there is an additional term $- \rho L^2\partial_L\Delta m^{2}$ in the above equation of motion. At the level of the equation of motion this is an effective contribution to the running of the anomalous dimension $\gamma$ that depends on the gradient of the rate of running in the gauge theory. At one loop in the gauge theory there is no such term and so we will neglect this term, effectively imposing the RG running of $\Delta m^2$ only at the level of the equations of motion. Since we are interested in theories that run from a trivial UV fixed point to an IR fixed point the dropped term would only influence the intermediate regime and then only for the smaller values of $N_f$ where the running is fast.  We have checked there is no qualitative change in the theory in the conformal window by including it. 

To find solutions for $L(\rho)$ and express the quark condensate in terms of the bare mass one needs to impose a regularity condition in the IR. The top-down D3-D7 system \cite{Karch:2002sh} has the IR condition $\partial_\rho L(0) = 0$ as that condition. However, this issue is more subtle in this model as we will show. The IR solutions do not satisfy $\partial_\rho L(0) = 0$ except in the conformal massless limit. We believe the reason for this is that the model does not include the backreaction to the quark flavour's mass (and condensate). Were the mass' backreaction to be included it would generate a small shift in the value of the dilaton at the scale of the mass as the flavours decouple from the QCD running. We would expect that variation in the geometry to accommodate a solution with $\partial_\rho L(0) = 0$. Rather than attempt the backreaction though we shall simply use an on-mass shell condition in the IR to terminate the RG flow. We discuss this issue in detail in the IR and UV.

In the full running theory at large energy scales, the running
of the anomalous dimension $\gamma$ is determined by the one loop QCD results. There is then a regime, around a scale we will call $\Lambda_{1}$, where the coupling is sufficiently
strong that the two loop contribution to the running of the coupling will become important and at scales somewhat below
this, the theory will approach an IR fixed point. A quark with large bare mass ($\gg \Lambda_1$) 
will only experience the high energy regime since it will be integrated from the theory at its mass scale which will be well above $\Lambda_1$. For quarks with very small bare mass ($\ll \Lambda_1$) their IR physics will be determined by the fixed point behaviour. It is therefore useful to study these two extreme regimes before looking at the full theory.

\subsection{\bf IR Fixed Point Behaviour of $c$}

In the IR of the conformal window $\alpha \rightarrow - b_0/b_1$, $\gamma_1$ becomes constant and hence $\Delta m^2$ is a non-zero constant. $\Delta m^2$ must lie in the regime $-1< \Delta m^2 <0$ for the theory to be stable and remain conformal in the IR without the chiral condensate forming.   Let us first for simplicity consider the theory that lives at the fixed point at all scales and so has no running of the coupling or $\gamma$.

The solutions of the RG flow equation (\ref{embedeqn}) are of the form
\begin{equation}\label{sol}
L = {m_{FP} \over \rho^\gamma} + {c_{FP} \over \rho^{2-\gamma}}, \hspace{1cm} \gamma(\gamma-2) = \Delta m^2 \end{equation}
here $m_{FP}, c_{FP}$ are interpreted as being the operator source combination for the operator $\bar{q} q$ but
of course in this theory they have dimensions $1+\gamma$ and $3 - \gamma$.

To extract the chiral condensate we substitute the solution back into the action (\ref{daq}), integrate over $\rho$ upto a cut off $\Lambda_{UV}$,  and compute ${1 \over Z} {d Z \over dm_{FP}}|_{m_{FP}}$. We find
\begin{equation} \begin{array}{ccl} \langle \bar{q} q \rangle_{FP} & = & {(\Delta m^2 + \gamma^2) \over (1 - \gamma) } 
m_{FP} \Lambda^{2-2 \gamma}_{UV}  \\ &&\\
&&+ ~~2 (\Delta m^2 + \gamma(2-\gamma) ) c_{FP} \ln \Lambda_{UV} \end{array}\end{equation}
The first term is the expected UV divergence in the condensate in the presence of a mass - the mass and condensate share the same symmetry properties and the dimension is then made up with the UV cut off scale. Since the condensate has dimension $3-\gamma$ and $m_{FP}$ dimension $1+\gamma$ the power of $\Lambda_{UV}$ is the correct one to match this dimensional analysis. This is already a sign that the model correctly describes scaling dimensions.  The second term is, upto log renormalization, a constant times the parameter $c_{FP}$ - in the $m_{FP}=0$ limit $c_{FP}$ is therefore proportional to the condensate. We will study $c_{FP}$'s scaling behaviour shortly.  

To find solutions for $L(\rho)$ and express $c_{FP}$ in terms of $m_{FP}$ one needs to impose a regularity condition in the IR.  The solutions in (\ref{sol})
clearly do not satisfy $\partial_\rho L(0) = 0$ except in the conformal $m=c=0$ limit. As we discussed this is
most likely a failure of the model to include the back reaction of the quark decoupling on the background. We will
rectify this by choosing a suitable boundary condition.
A sensible first guess for the IR  boundary condition is
\begin{equation}\label{bca}  L (\rho=L_0) = L_0, \hspace{1cm}  L'(\rho=L_0)=0. \end{equation}
This IR condition is similar to that from top down models but imposed at the RG scale where the flow becomes ``on-mass-shell". Here we are treating $L(\rho)$ as a constituent quark mass at each scale $\rho$. We then find 
\begin{equation}
m_{FP} = \left({\gamma-2 \over 2\gamma -2} \right) L_0^{1+\gamma} 
\end{equation}
and
\begin{equation}\label{qqir}
c_{FP} = {\gamma \over 2\gamma-2} \left( {2\gamma -2 \over \gamma-2} \right)^{3-\gamma \over 1+\gamma} m_{FP}^{3-\gamma \over 1+\gamma} 
\end{equation}
This shows analytically that the model obeys the ``hyper-scaling" relation one would expect at the conformal fixed point. The condensate has dimension $3-\gamma$ and the mass dimension $1+\gamma$. Since $m_{FP}$ is the only intrinsic scale 
$c_{FP} \sim m_{FP}^{3-\gamma/1+ \gamma}$ is ensured. In the full theory with a running coupling relations of this form will hold in any regime where $\gamma$ is running slowly with the $c$ and $m$ parameters those appropriate to that energy regime.

The boundary condition $L'(\rho=L_0)=0$ is not crucial to obtain the hyper-scaling relations since the relative dimensions of $m_{FP}$ and $\bar{q}q_{FP}$ are fixed in the holographic model. Instead the choice of this boundary condition fixes
the proportionality constant between $\bar{q}q_{FP}$ and $m_{FP}^{3-\gamma/1+ \gamma}$. Given that there is some freedom in this choice of boundary condition, we will not be predicting this value -  for this reason in our numerics we will choose a  boundary condition to set the proportionality constant to unity in all cases. That is, we will assume at the IR boundary the solution is of the form
\begin{equation} \label{irassume}L=\frac{m_{FP}}{\rho^\gamma}+\frac{m_{FP}^\frac{3-\gamma}{1+\gamma}}{\rho^{2-\gamma}}.
\end{equation}
and hence use the boundary conditions
\begin{equation}\label{bc}\begin{array}{c}  L (\rho)|_{L_0} = L_0,\\ \\L'(\rho)|_{L_0}=-\frac{\gamma m_{FP}}{L_0^{\gamma+1}}+\frac{\gamma-2}{L_0^{3-\gamma}}m_{FP}^{\frac{3-\gamma}{1+\gamma}}.\end{array}
\end{equation}
Note here that the value of $\gamma$ used in the initial condition is that determined by (\ref{dmsq3}) (and the discussion below) evaluated at the scale $\mu = \sqrt{2} L_0$.

\subsection{\bf Large $m$ limit}

If we now consider asymptotically free theories that lie at $\alpha<\alpha_*$ in the UV then the far UV running of $\Delta m^2$ is controlled by the one loop perturbative running coupling. Theories where the $L$ profile lives at large values of $r=\sqrt{L^2+\rho^2}$ will see only this behaviour ie we can extract the large quark mass behaviour from this limit.
The embedding equation is
\begin{equation}
\partial_\rho [ \rho^3 \partial_\rho L] - {\rho \kappa \over \ln \rho / \Lambda_1}   L = 0,
\end{equation}
where $\Lambda_1$ is the one loop running scale and $\kappa$ is a constant which, at the one loop level, can be shown to take the form $\kappa=-\frac{3}{2}\frac{N_c^2-1}{N_c\pi b_0}$.  The solution thus has the behaviour
\begin{equation} \begin{array}{ccl} \label{solUV}
L &=&  m(\rho) + {c(\rho) \over \rho^2}  \\ && \\
&= & 
 {\nu_{UV} \over (\ln \rho / \Lambda_1)^k} + {c_{UV} \over \rho^2} (\ln (\rho / \Lambda_1))^k, \hspace{1cm} k=-{\kappa \over 2}. \end{array}\end{equation}
 To obtain the bare quark mass one simply extracts the non-normalizable term of the solution at some fixed UV scale (we choose $\rho=e^{500}$ for the numerical work below) - this is what we will refer to as $m_\text{bare} $ in plots that follow.

Applying the simple boundary conditions
\begin{equation}  L (\rho=L_0) = L_0, \hspace{1cm}  L'(\rho=L_0)=0, \end{equation}
gives
\begin{equation}
\nu_{UV} = - {2 c_{UV} \over k L_0^2} \left(\ln { L_0 \over \Lambda_1}\right)^{2k+1},
\end{equation}
and
\begin{equation}
c_{UV} = -{k \over 2 \left(\ln {L_0 \over \Lambda_1}\right)^{4k+1} } \nu_{UV}^3. 
\end{equation}
This shows that $c_{UV} \sim \nu_{UV}^3$ in the UV upto a logarithmic renormalization.  The model is again correctly determining the scaling relations between the mass and condensate though. 

We assumed that $L'(L_0)=0$ here so that we could display the scaling behaviours analytically. In our numerical work we will use the boundary condition in (\ref{bc}) which sets $c_{UV}=\nu_{UV}^3$ in the IR for large quark masses also.

\begin{figure}[]
\centering
\includegraphics[width=6.5cm]{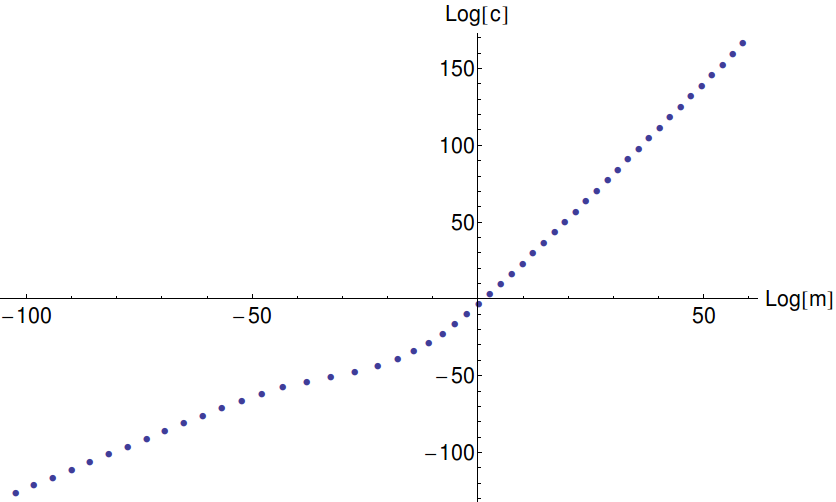} \includegraphics[width=6.5cm]{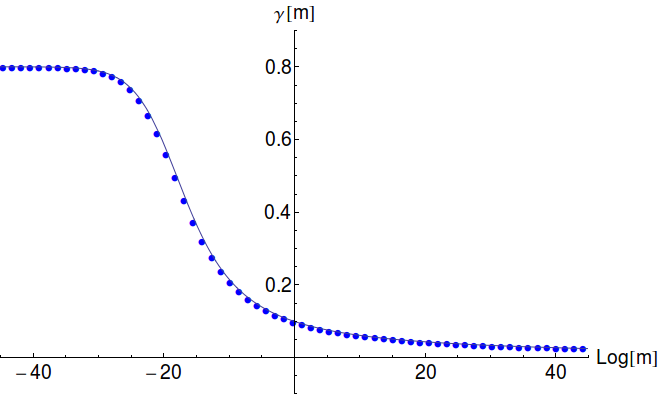} \includegraphics[width=6.5cm]{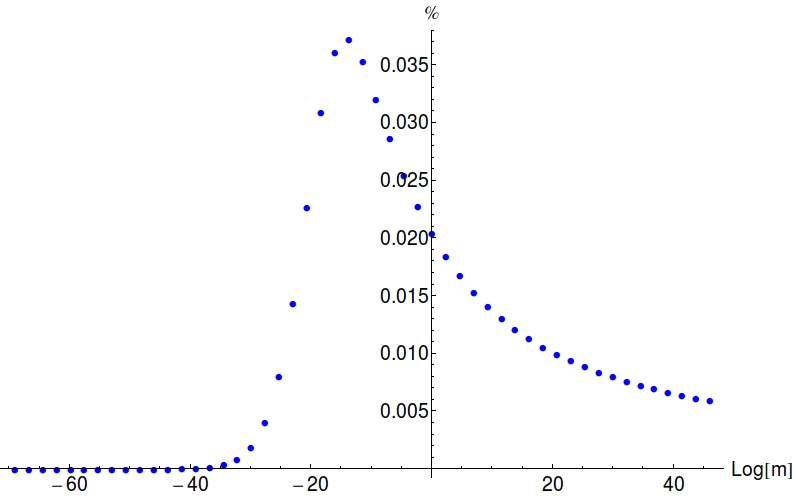}
\caption{Plots for the theory $N_c=3$ and $N_f=12$: \textbf{a}) Log $c$ against Log $m_\text{bare}$  ~~~ \textbf{b}) numerical points for $\gamma$ extracted from $b=\frac{3-\gamma}{1+ \gamma}$ against Log $m_\text{bare}$ (the solid line is the input $\gamma$ from the two loop running \textbf{c)} percentage difference between the extracted form of $\gamma$ and the input form (solid line in b)) }
\label{nf12qq}
\end{figure}

\subsection{Numerical Solutions for the Full Running Theory}

We have seen that the model correctly describes scaling dimensions in the IR and UV fixed point regimes. 
The transition between these fixed points is more model dependent but also of more interest for lattice simulations where one would be interested in an estimate of how quickly the IR scaling behaviour is likely to set in. We can see what results this model gives by numerically solving for $c$ and $m$ as a function of RG scale with the full two loop running implemented. 

We first discuss results for $N_c=3$ and $N_f=12$ as an example. This model lies close to the lower edge of the  conformal window ($N_f^c< N_f<11N_c/2$) which does not exhibit spontaneous chiral symmetry breaking ($\gamma<1$). Specifically, it displays an IR fixed point value for $\gamma$ of $\gamma^*$=0.8  (note this fixed point value follows from the one and two loop QCD beta-functions).
We proceed by solving (\ref{embedeqn}) subject to the IR boundary condition (\ref{bc}). Then at each value of $\rho$ we fit $L,L'$ and $L''$ to the functional form 
\begin{equation}
L = {m \over \rho^\gamma} + {c \over \rho^{2-\gamma}}
\end{equation}
to extract an estimate of the running mass, condensate and $\gamma$. Note here $m$ is a parameter that in the UV has dimension 3
and displays logarithmic running consistent with the discussion of (\ref{solUV}) whilst in the IR it runs to the IR dimension $1+\gamma$ source. 
This ansatz for the fitting is sound in the UV and IR fixed point regimes and will likely be good locally in slowly running regimes but is neccessarily approximate.  

Let us first evaluate the condensate at the deepest IR point for each value of quark mass ie $L_0$ for each flow. We have fixed $L'$ at this point assuming that the solution takes the form in (\ref{irassume}) so in the IR and UV fixed point regimes (ie at low and high quark mass) we expect the numerical solution to match that form precisely. In the intermediate regime where $\gamma$ is running, the form in (\ref{irassume}) is only approximate. The numerical solutions for the quark condensate parameter $c$ against the quark mass are displayed in Fig \ref{nf12qq}a).  The plot shows clear UV and IR scaling regimes where $c \sim m^b$ with a transition period between.

In Fig \ref{nf12qq}b) the value of $\gamma$ extracted from $b$ is plotted over the input form of $\gamma$ as discussed below Eq (\ref{dmsq3}). Assuming $b$ takes the form $b=\frac{3-\gamma}{1+\gamma}$, one should expect to return the input value of $\gamma$, since the IR regularity condition imposes that $c=m^\frac{3-\gamma}{1+\gamma}$ and we are evaluating $c$ at the IR boundary. It is clear from Fig \ref{nf12qq}b) that the extracted $\gamma$ does indeed agree very well with the input form bar marginal discrepancies in the regime of steepest running.
The extent of the deviation in this intermediate regime can be seen more clearly in Fig \ref{nf12qq}c) as a percentage difference from the input form. Clearly the ansatz (\ref{sol}) works well at all scales.
The slight deviation between the input and output $\gamma$, which reflects the additional scale $\Lambda_1$ from the running, seems to persist for several decades of energy on either side of the strongest running regime in this model. Such behaviour, if true of the full theory, would further complicate lattice studies of such theories by requiring a very large box size to include both the UV and IR fixed point behaviours.

\begin{figure}[]
\centering
\includegraphics[width=6.5cm]{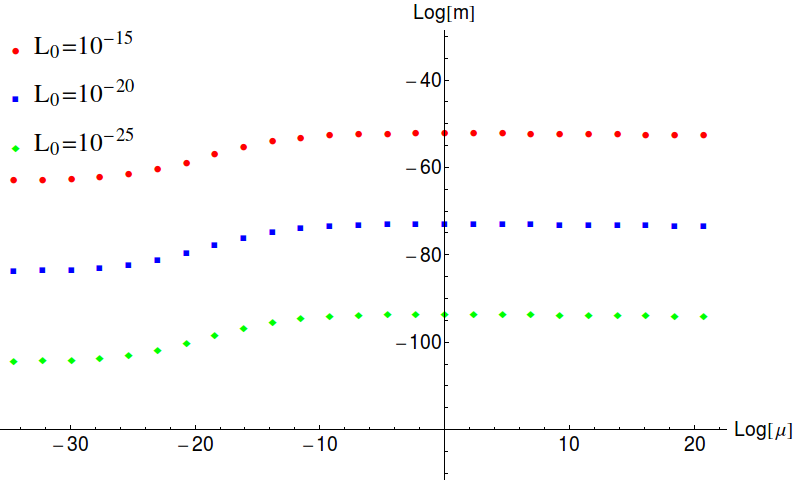} \includegraphics[width=6.5cm]{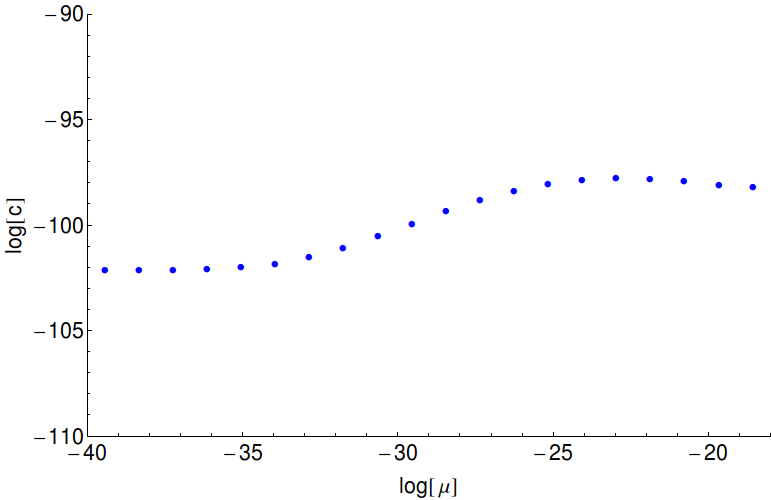} 
\caption{Plots for the running of mass and condensate at $N_c=3$, $N_f=12$: \textbf{a)} The running of the mass shown against RG scale for different values of $L_0$ \textbf{b)} The running of the condensate parameter $c$ against RG scale $\mu$ at $L_0=10^{-20}$. }
\label{nf12massrun}
\end{figure}

\begin{figure}[]
\centering
\includegraphics[width=6.5cm]{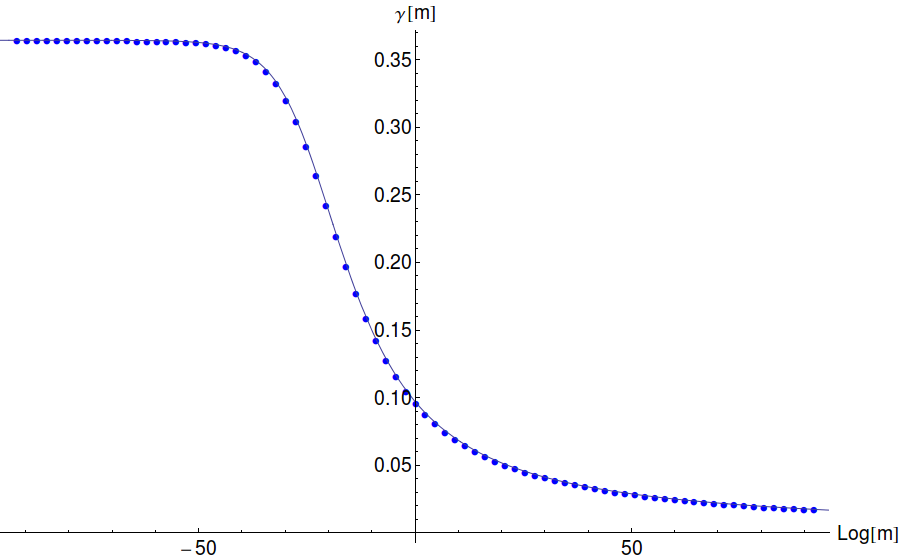} \includegraphics[width=6.5cm]{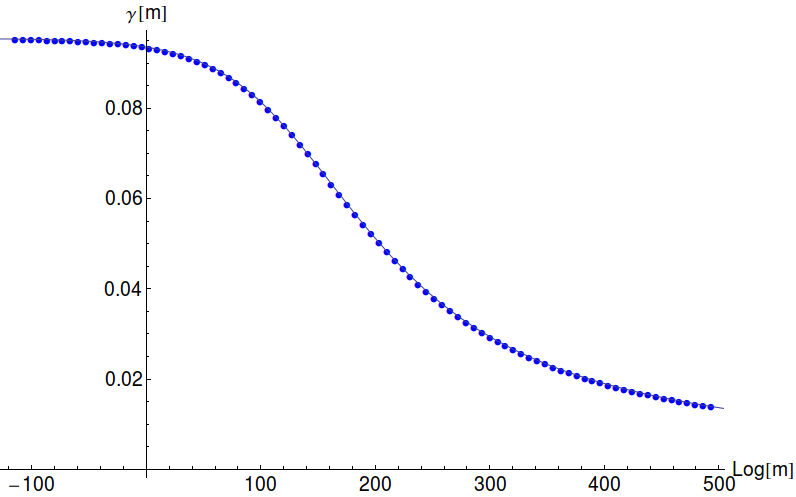} 
\caption{$N_c=3$: \textbf{a)} $\gamma$ versus $m_\text{bare}$ from $c$  for $N_f=13$ \textbf{b)} $\gamma$ versus $m_\text{bare}$ from $c$ for $N_f=15 $ }
\label{1315qq}
\end{figure}
\bigskip

It is now interesting to follow the RG flow of the quark mass and condensate parameter $c$ for a particular choice of IR mass. In Fig \ref{nf12massrun}a)  we display this running for a quark mass that lies in the IR conformal regime. It is important to stress again that the dimension of these objects changes along the flow - in the case of the mass for example from $1+\gamma$ to 1 between the IR and UV. In the deep IR there is no significant running of the  source and this is in agreement with the conformal behaviour of $\gamma$ in this region. As the running scale is increased we see a step-up in the quark mass value until it plateaus in the UV  to logarithmic running. The RG flow of the quark condensate seen in \ref{nf12massrun}b) also follows a similar behaviour, stepping up from a fixed point value in the IR as we increase the RG scale. The large step in the value of these parameters is expected. At the scale $\Lambda_1$ their dimension is changing significantly and the scale $\Lambda_1$ is the only available scale with which to adjust their dimension and it is much bigger than the initial values of $m,c$ - large renormalization effects are expected. This argument is essentially the one used to predict that walking gauge dynamics \cite{Holdom:1981rm} will have a large UV condensate and these plots support that logic.    An alternative presentation of this data would be to define a mass term of fixed dimension 1 all along the flow by
$m(\rho) = m / \rho^\gamma$ - this quantity would again match the usual intuition in the UV but scale as $\rho^{-\gamma}$ in the IR regime.

The behaviour for other values of $N_f$ in the conformal window are very similar in spirit to the $N_f=12$ case we have looked at in detail. To summarize the other cases we simply produce the plot of $\gamma$ extracted from the fit of the form $c \sim m^{(3-\gamma)/( 1 + \gamma)}$  against quark mass overlaid on the input $\gamma$ function from the two loop running. We show results for the cases $N_f=13$ and $N_f=15$ in Fig \ref{1315qq}. These plots indicate that the aforementioned discrepancy in the regime of strongest running becomes increasingly less dominant at higher values of $N_f$. This trait encapsulates the idea that as the number of flavours increases, the fixed point value of $\gamma$ drops and the rate of running slows causing the IR fixed point behaviour to extend further away from $\mu=0$.

\section{Bound State Masses} 

So far our analysis has consisted of checking that the vacuum configuration of the model is consistent with na\"ive scaling arguments. One of the powers of holographic models is that these relations are inbuilt.  We now turn to computing the physical parameters, the masses of the bound states and their decay constants. These parameters are true predictions of the model now that the dynamics has been included through the running scalar mass and the condensate fixed by the IR boundary condition. 

\subsection{Linearized Fluctuations} 

The scalar $\bar{q}q$ ($\sigma$) mesons are described by linearized fluctuations of $L$ about its vacuum configuration, $L_v$. We look
for space-time dependent excitations,  ie $|X| = L_v + \delta(\rho) e^{-i q.x}$,  $q^2=-M_\sigma^2$. The equation of motion for $\delta$ is, linearizing (\ref{embedeqn}),
\begin{equation} \label{deleom} \begin{array}{c}\partial_\rho( \rho^3 \delta' ) - \Delta m^2 \rho \delta -   \rho L_v \delta \left. \frac{\partial \Delta m^2}{\partial L} \right|_{L_v} \\ \\ 
+ M_\sigma^2 R^4 \frac{\rho^3}{(L_v^2 + \rho^2)^2} \delta  = 0\,. \end{array} \end{equation}
We seek solutions with, in the UV, asymptotics of $\delta=\rho^{-2}$ and with $\partial_\rho\delta|_{L_0}=0$ in the IR, giving a discrete meson spectrum. Note that the distinction between this IR boundary condition and that of the normalizable mode in (\ref{bc}) is negligible in the spectrum obtained (of order 1 part in 10$^{5}$). Recalling previous discussion of the $\partial_L\Delta m^2$ term, we elect to ignore it since it has negligible effects on the  spectrum.

\begin{figure}[]
\centering
\includegraphics[width=6.5cm]{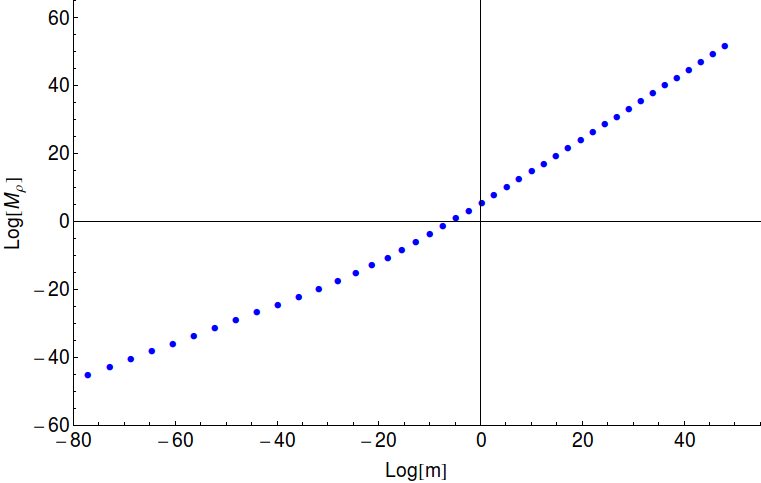} \includegraphics[width=6.5cm]{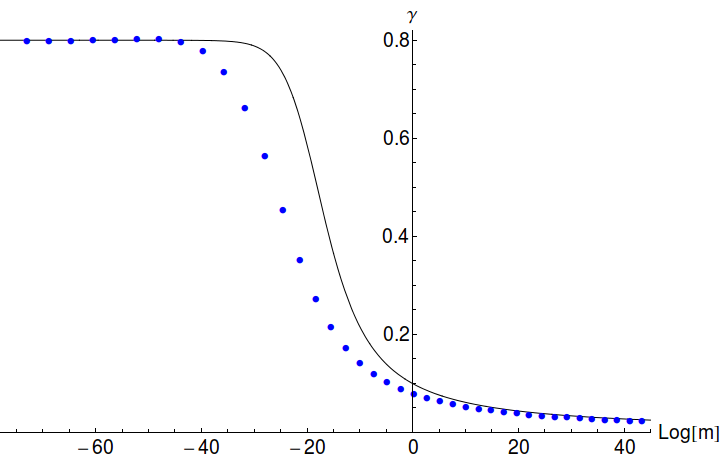}  \includegraphics[width=6.5cm]{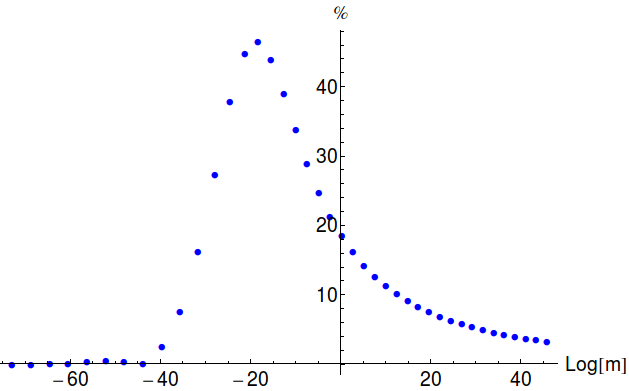} 
\caption{$N_c=3$, $N_f=12$: \textbf{a)} $\rho$-meson mass against quark mass   \textbf{b)} Extracted value $\gamma$ versus $m_\text{bare}$ from $\rho$-meson mass spectum. The solid line shows the holographic input of $\gamma$ from the two-loop running \textbf{c)} The percentage difference seen between the input $\gamma$ running and the extracted $\gamma$ running}
\label{nf12mesonr}
\end{figure}

\begin{figure}[]
\centering
\includegraphics[width=6.5cm]{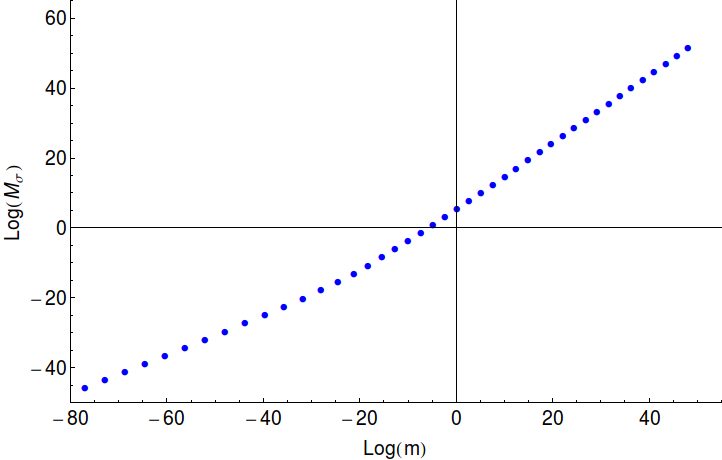} \includegraphics[width=6.5cm]{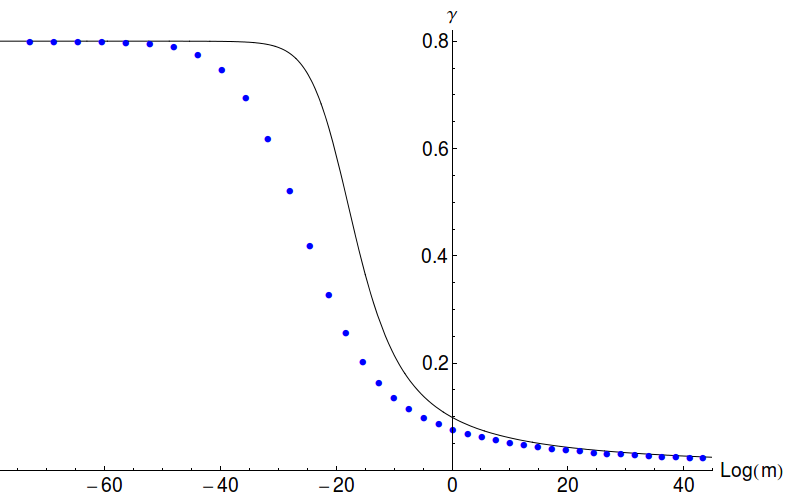} 
\caption{$N_c=3$, $N_f=12$:\textbf{a)} $\sigma$-meson mass against quark mass  \textbf{b)} Extracted value $\gamma$ versus $m_\text{bare}$ from $\sigma$-meson mass spectum. The solid line shows the holographic input of $\gamma$ from the two-loop running.}
\label{nf12mesons}
\end{figure}

We must normalize $\delta$ so that the kinetic term of the $\sigma$ meson is canonical i.e.
\begin{equation} \int d \rho {\rho^3 \over (\rho^2 + L_v^2)^2} \delta^2 = 1\,. \end{equation}

The scalar meson decay constant can be found using the solutions for the normalizable and non-normalizable wave functions. We concentrate on the action term (after integration by parts)
\begin{equation}
S   = \int d^4x~ d \rho ~~ \partial_\rho (- \rho^3 \partial_\rho L) L\,.
\end{equation}
We substitute in the normalized solution $\delta$ and the external non-normalizable scalar function $K_S$
at $q^2=0$ with normalization $N_S$ to obtain the dimension one decay constant $f_\sigma$ as 
\begin{equation}
f_\sigma^2 =  \int d \rho \partial_\rho (- \rho^3 \partial_\rho \delta)  K_S(q^2=0)\,.
\end{equation}

The vector ($\rho$) meson spectrum is determined from the normalizable solution of the equation of motion for the spatial pieces of the vector gauge field $V_{\mu \perp} = \epsilon^\mu V(\rho) e^{-i q.x}$ with $q^2=-M^2$. The appropriate equation is
\begin{equation} \label{vv}  \partial_\rho \left[ \rho^3 \partial_\rho V \right] + {\rho^3 M^2 \over (L_v^2 + \rho^2)^2} V = 0\,. \end{equation}
We again impose $\partial_\rho V|_{L_0}=0$ in the IR and require in the UV that $V\sim c/\rho^2$. To fix $c$ we normalize the wave functions such that the vector meson kinetic term is canonical
\begin{equation}  \int d \rho {\rho^3  \over  (\rho^2 + L_v^2)^2} V^2 = 1\,. \end{equation}

The vector meson decay constant is given by substituting the solution back into the action and determining the coupling to an external $q^2=0$ vector current with wave function $K_V$. We have for the dimension one $f_V$
\begin{equation} f_V^2 = \int d \rho  \partial_\rho \left[- \rho^3 \partial_\rho V\right] K_V(q^2=0)\,.
\label{rhodecay}
\end{equation} 
The pion mass spectrum is identified by assuming a space-time dependent phase $\pi^a(x)$ of the AdS-scalar $X$ describing the $\bar{q}q$ degree of freedom, i.e $X=L(\rho)\exp(2i\pi^a(x)T^a)$. The equation of motion of the pion field is then,
\begin{equation}
\partial_\rho\left(\rho^3L_v^2\partial_\rho\pi^a\right)+M_\pi^2\frac{\rho^3L_v^2}{(\rho^2+L_v^2)^2}\pi^a=0.
\label{pionfield}
\end{equation}
Again, we impose at the IR boundary that $\partial_\rho\pi^a|_{L_0}=0$.

\begin{figure}[]
\centering
\includegraphics[width=6.5cm]{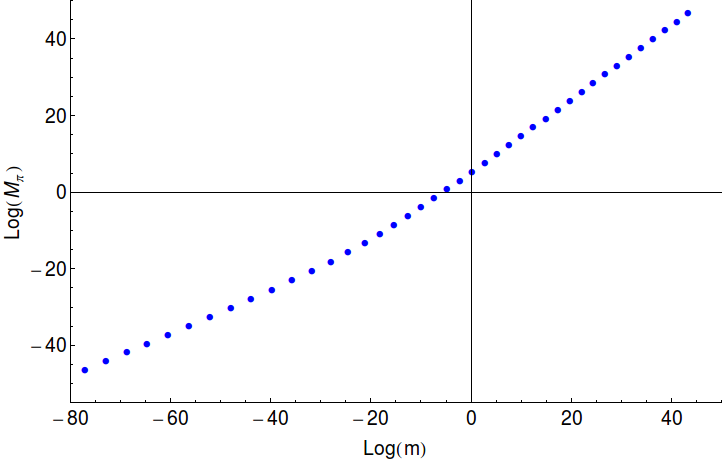} \includegraphics[width=6.5cm]{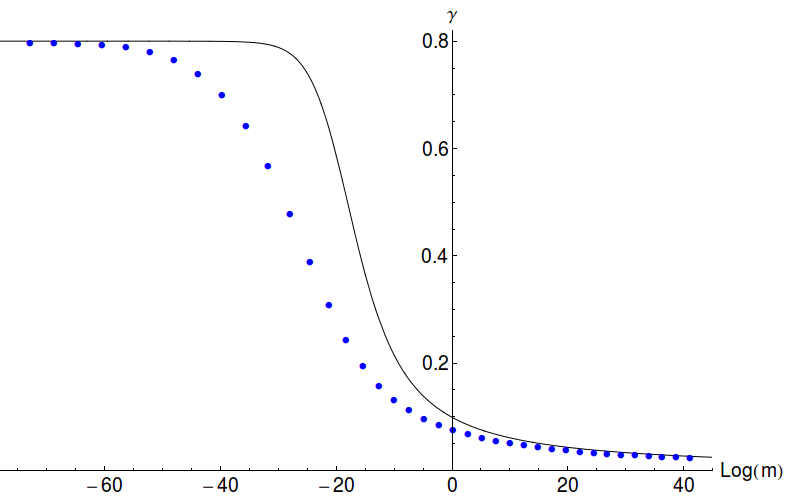} 
\caption{$N_c=3$, $N_f=12$:\textbf{a)} $\pi$-meson mass against bare quark mass  \textbf{b)} Extracted value $\gamma$ versus $m_\text{bare}$ from $\pi$-meson mass spectum. The solid line shows the holographic input of $\gamma$ from the two-loop running.}
\label{nf12mesonp}
\end{figure}

\subsection{Bound States of the $N_c=3, N_f=12$ Theory}

We again focus in detail on the $N_c=3, N_f=12$ theory with $\gamma^* \simeq 0.8$. Using the formalism described we compute the $\rho, \pi$ and $\sigma$ meson masses as a function of quark mass. Hyper-scaling arguments lead to the expectation that in a fixed point theory the meson mass will scale as $m_\text{bare}^{1/1+\gamma}$. In the UV $\gamma=0$ whilst in the IR $\gamma=0.8$. In Figs \ref{nf12mesonr}a), \ref{nf12mesons}a) and \ref{nf12mesonp}a), we plot the dependence on the $\rho$-mass, $\sigma$-mass and the $\pi$-mass respectively, against the bare quark mass $m_\text{bare}$. Note here we define the bare quark mass as the running quark mass evaluated at a very high UV scale of $\rho = e^{500}$. In Figs \ref{nf12mesonr}b), \ref{nf12mesons}b) and \ref{nf12mesonp}b), we plot $\gamma$ extracted from the hyperscaling relation, again as a function of the quark mass, and show the comparison to the input running of $\gamma$. As for the quark condensate we see excellent agreement with the hyperscaling relation in the UV and IR regimes but a discrepancy in the intermediate running region. In the central region the discrepancy again reflects the presence of the second scale $\Lambda_1$ in the running coupling. The deviations from the na\"ive IR and UV fixed point values seem to persist in the meson masses over a slightly wider running period than in the input $\gamma$. The percentage deviation in $\gamma$  extracted from the $\rho$ mass and the initial two-loop $\gamma$ input is shown in Fig \ref{nf12mesonr}c). In the regime of strongest running, the disagreement is found to be as much as 47\%. 

\begin{figure}[]
\centering
\includegraphics[width=6.5cm]{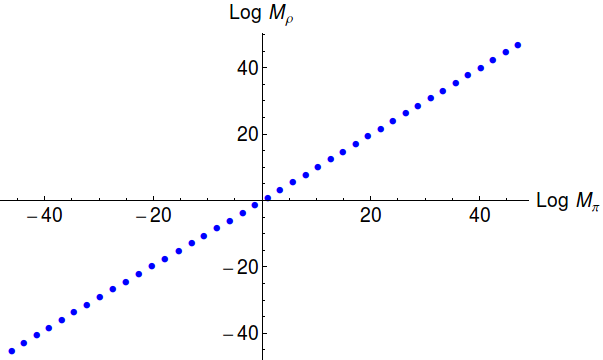} \includegraphics[width=6.5cm]{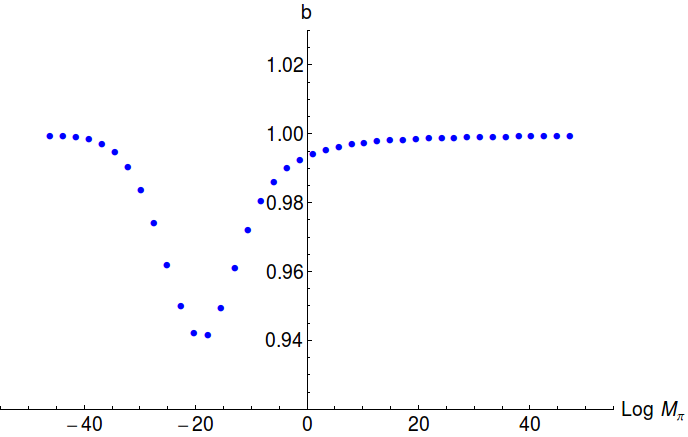} 
\caption{$N_c=3$, $N_f=12$: \textbf{a}) $\rho$-mass versus $\pi$-mass   \textbf{b}) $b$ versus $M_\pi$, where we've assumed $M_\rho\propto M_\pi^b$ }
\label{mpimrho12}
\end{figure}
\begin{figure}[]
\centering
\includegraphics[width=6.5cm]{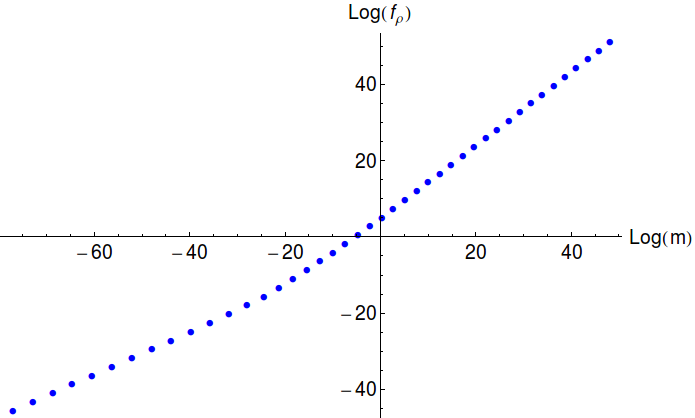} \includegraphics[width=6.5cm]{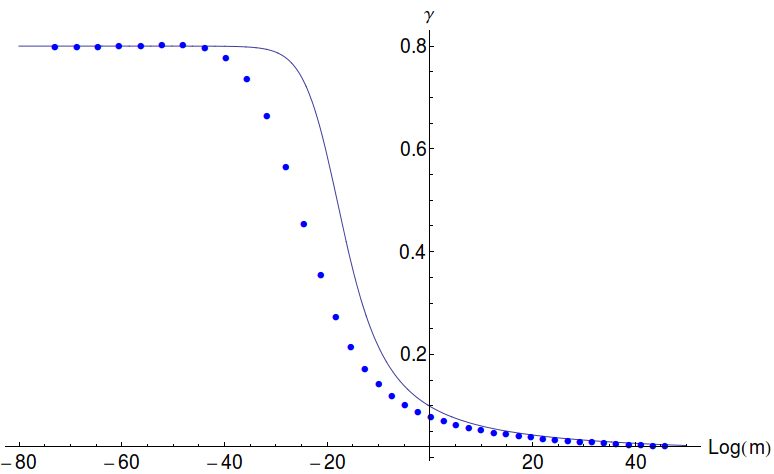}
\caption{$N_c=3$, $N_f=12$:\textbf{a)} $f_\rho$ versus the bare quark mass \textbf{b)} the extracted $\gamma$ versus $m_\text{bare}$ }
\label{f12r}
\end{figure}
\begin{figure}[]
\centering
\includegraphics[width=6.5cm]{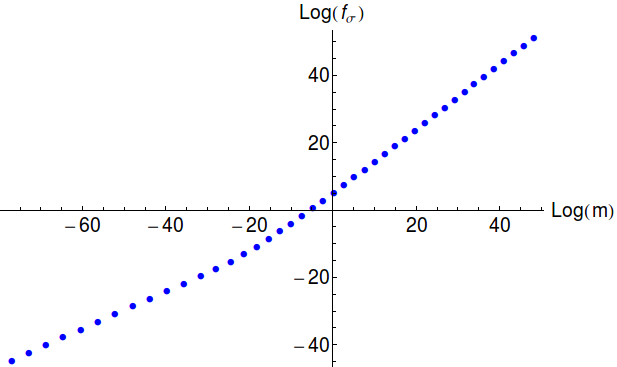} \includegraphics[width=6.5cm]{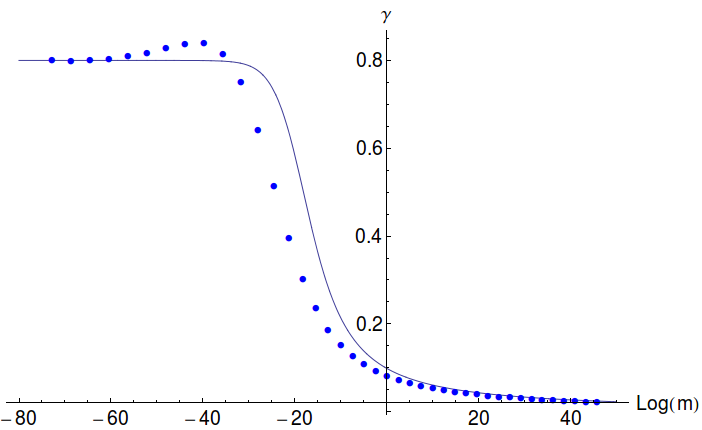}  
\caption{$N_c=3$, $N_f=12$:\textbf{a)}$f_\sigma$ versus the bare quark mass \textbf{b)} the extracted $\gamma$ versus $m_\text{bare}$  }
\label{f12s}
\end{figure}

Another interesting plot is to remove the unphysical quark mass and directly plot $M_\rho$ versus $M_\pi$. Here we expect at a fixed point that $M_\rho \propto M_\pi^b$ with $b=1$. In Fig \ref{mpimrho12}  we plot these masses against each other and the extracted value of $b$ against $M_\pi$.  We indeed see the expected proportionality between the masses in the fixed point regimes as well as the deviation in the running regime between these, a telltale sign of the running scale $\Lambda$ entering the relation. Here the deviations from the fixed point scaling is only of order 5$\%$.

Finally, we can compute the decay constants $f_\rho$ and $f_\sigma$ and plot them against the quark mass, see Figs \ref{f12r}a) and \ref{f12s}a). Once again, we extract $\gamma$, assuming a power law relationship $f_{\rho,\sigma}\propto m_{\rm bare}^{1 \over 1 + \gamma}$ and plot the results in Figs \ref{f12r}b) and \ref{f12s}b). They show similar behaviour to the meson masses.

\subsection{$N_c=3$, $N_f=13,15$ Mesons}

\begin{figure}[]
\centering
\includegraphics[width=6.5cm]{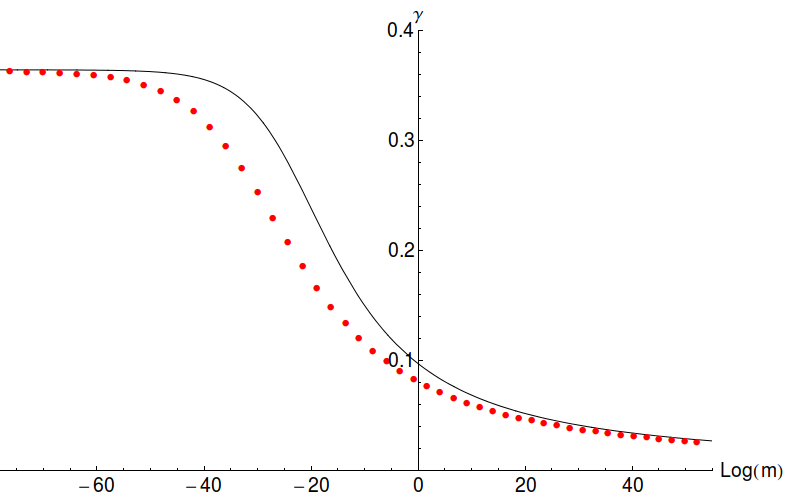} \includegraphics[width=6.5cm]{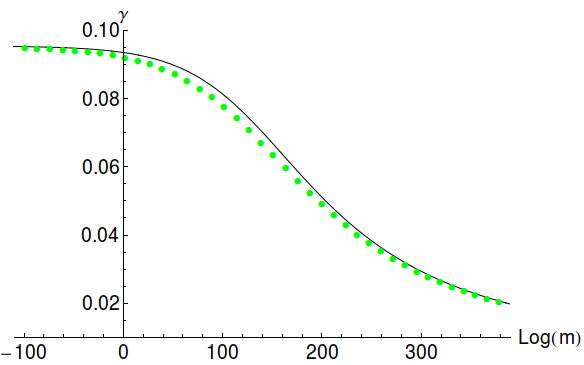}
\caption{$N_c=3$: \textbf{a)} $\gamma$ extracted from $M_\rho$ against the bare quark mass at $N_f=13$, $N_c=3$ \textbf{b)} The same for $N_f=15$.}
\label{mrho13/5}
\end{figure}

For completion we have also computed the mesonic variables at $N_f=13$ and $N_f=15$ in the $N_c=3$ theory, so that we can test this model across a large span of the conformal window. We begin, as before, by computing the mass spectra of the $\rho$- and $\sigma$-mesons as a function of the quark mass and extract the corresponding $\gamma$, which can be seen in Fig \ref{mrho13/5}a) for $N_f=13$  and Fig \ref{mrho13/5}b) for $N_f=15$. A similar behaviour to that at $N_f=12$ is observed with the clear IR and UV scaling regimes of $M_{\rho}\propto m_{\rm bare}^{\frac{1}{1+\gamma^*}}$ and $M_\rho\propto m_{\rm bare}$ respectively. We see the deviation from the input $\gamma$ running in the central region where the running is strongest. However, as $N_f$ is increased away from $N_f^c$, the IR fixed point value, $\gamma^*$, decreases  thus reducing the rate of the running with RG scale so the deviation in $\gamma$ becomes less and less. It is most evident for the case $N_f=15$ in Fig \ref{mrho13/5}b), that not only does the discrepancy between the input $\gamma$ and the extracted $\gamma$ become less pronounced with increased $N_f$ (at most only $\sim4.8$\% difference compared to $47\%$ for $N_f=12$), but that the conformal IR fixed point behaviour gets `pushed' further from the scale $\Lambda_1$  so that one must overview many decades of RG scale to observe significant running. 

\begin{figure}[]
\centering
\includegraphics[width=6.5cm]{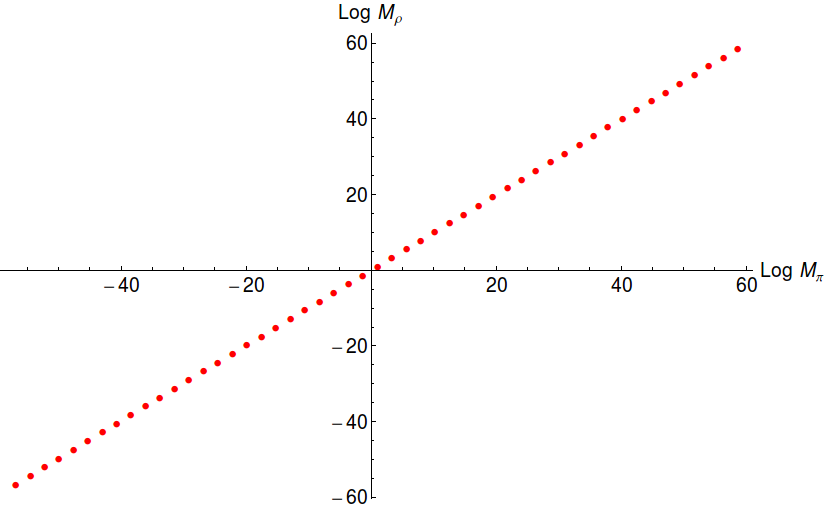} \includegraphics[width=6.5cm]{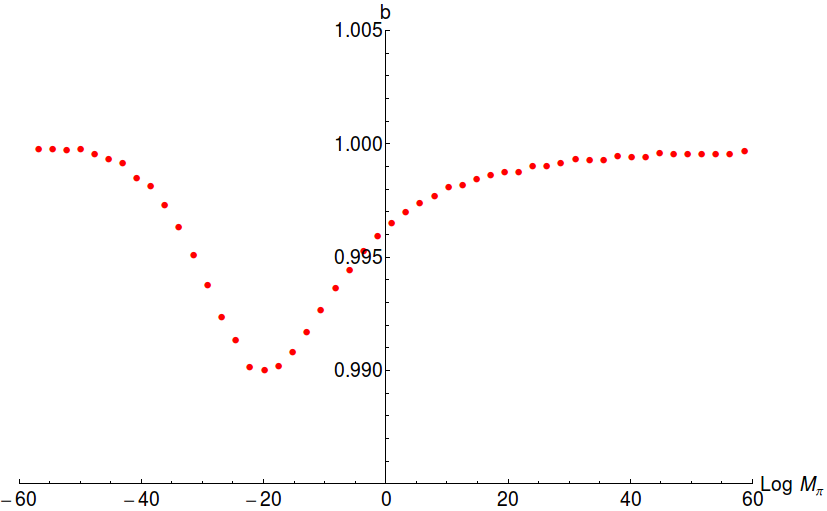} \includegraphics[width=6.5cm]{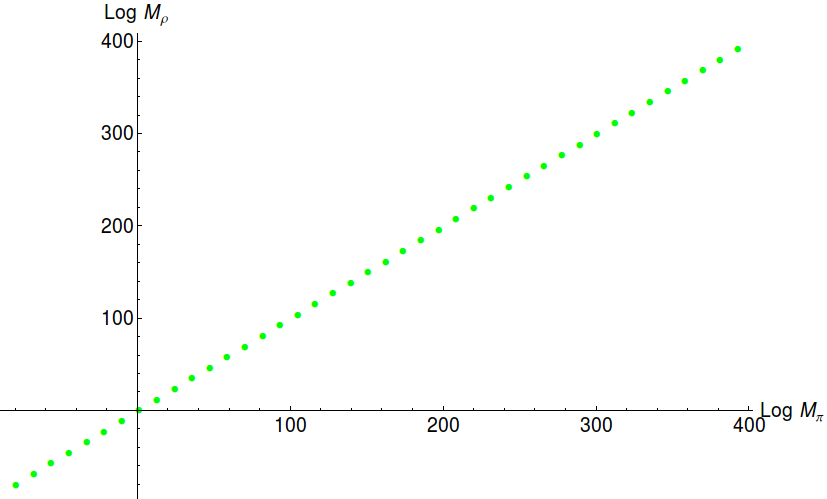} \includegraphics[width=6.5cm]{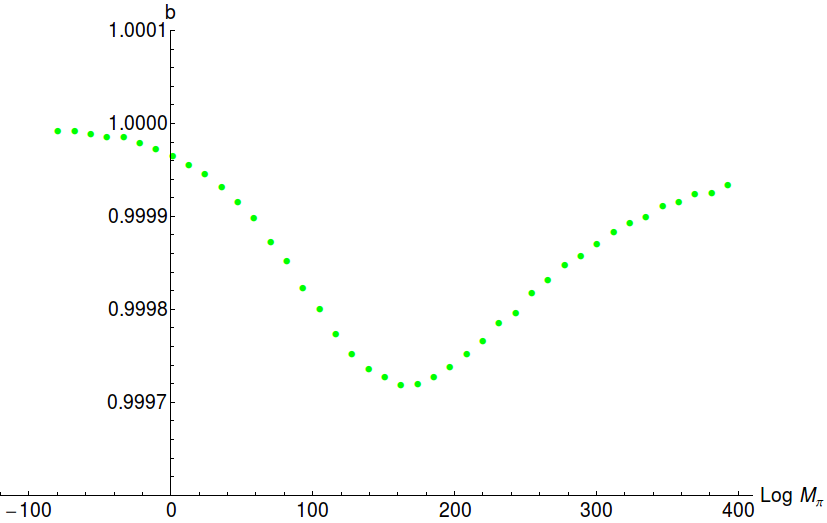}
\caption{$N_c=3$: \textbf{a)} $M_\rho$ versus $M_\pi$ for $N_f=13$ \textbf{b)} extracted $\gamma$ for $N_f=13$ \textbf{c)} $M_\rho$ versus $M_\pi$ for $N_f=15$ \textbf{d)} extracted $\gamma$ for $N_f=15$}
\label{mrmp1315}
\end{figure}

Next we turn again to plots of $M_\rho$ versus $M_\pi$ which remove the unphysical mass parameter $m_{\rm bare}$, see Figs \ref{mrmp1315}a) and \ref{mrmp1315}c). In each of the cases $N_f=13$ and $N_f=15$, the linear relationship $M_\rho\propto M_\pi$, expected in the IR and UV regions, is clearly observed and only by examining the exponent, $b$, of an assumed $M_\rho\propto M_\pi^b$ relationship do we  notice the discrepancy attributed to the additional running scale $\Lambda_1$; see Figs  \ref{mrmp1315}b) and \ref{mrmp1315}d). Once more we observe that an increase in the number of flavours leads to an extended IR fixed point region and a reduction in the rate of running of the anomalous dimension with RG scale. Fig \ref{mrmp1315}d) showing $b$ versus $M_\pi$ at $N_f=15$ provides a prime example of such an observation - the  greatest difference between the extracted value of $b$ and the linear behaviour ($b=1$) is only of the order of 0.03\%.

\bigskip

\section{Summary}

Dynamic AdS/QCD is a holographic model of QCD with variable $N_c$ and $N_f$ which predicts the dynamics of chiral symmetry breaking \cite{Alho:2013dka} and the meson spetrum . The gauge dynamics is input through the running of the anomalous dimension of the quark mass $\gamma$. Here we used the two loop perturbative result for $\alpha$ to parametrize that running. Chiral symmetry breaking occurs when the IR fixed point value of the anolmalous dimension $\gamma^*$ grows to one (corresponding to a critical value of the number of flavours $N_f^c \simeq 4 N_c$). The chiral symmetry breaking case of $N_f < N_f^c$ was studied in \cite{Alho:2013dka}. Here we have focused on the conformal window of SU$(N_c)$ gauge theories with $N_f > N_f^c$ which do not exhibit spontaneous chiral symmetry breaking (s$\chi$SB) in the massless limit. The massless theory is conformal so we have studied the inclusion of a quark mass. 
The model does not directly include back-reaction on the glue dynamics due to the presence of the quark mass 
so we have adopted an on mass shell boundary condition where we shoot out from RG scales of order the constituent quark mass. We have shown carefully that the model correctly incorporates the scaling dimensions of the quark mass and condensate following the hyper-scaling power relations expected of the theory at the UV and IR fixed points. 

The main predictions of the model are the physical meson masses (of the $\rho, \pi$ and $\sigma$) and their decay 
constants. We have shown that the model reproduces the expected hyperscaling relation with respect to the bare (UV) quark mass, $M \sim m_\text{bare}^{1 \over 1 + \gamma}$, in the fixed point regimes. The model then allows us to study the deviations from this na\"ive scaling in the intermediate running regime between the fixed points characterized by the one-loop scale $\Lambda_1$. In this regime the theory has two scales and the simple power relations are expected to break down. We studied the SU(3) gauge theory with $N_f=12$ in detail. This theory has $\gamma^*\simeq 0.8$ and reasonably strong running - we indeed see substantial deviations from the simple power relation in the intermediate regime. The model also allows us to see that the regime in quark mass over which the simple power law behaviour will be absent is many orders of magntiude. The deviation from the relationship $M_\rho = M_\pi$ (expected in a one scale theory) is smaller (of order 5$\%$ in the power). This estimates the accuracy that would be needed in a lattice simulation on quantities such as $M_\rho$ to be certain that the fixed point had been achieved. For large $N_f=13,15$ the running is slower and the change in $\gamma$ from the UV to the IR smaller - the deviations from the simple powerlaw scaling are then smaller too (although spread over a larger energy regime). 
 
We have displayed results for $N_c=3$ and varying $N_f$ only. Whilst the true complete $N_c, N_f$ dependence is interesting, our model is underpinned by our ability to know the running of $\gamma$ in a particular theory. The use of the two loop running results is a decent parametrization but probably doesn't capture the true dynamics. For this reason we believe we have extracted the broad qualitative behaviours of the model on changing $\gamma^*$ and that these can be qualitatively brought across to other values of $N_c$. A true lattice simulation would be needed to pin the physics down more precisely. Our results show though that the physics of these theories is spread over quite a range of orders of magnitude if one wishes to flow from the UV perturbative fixed point to the IR fixed point regime. This remains a big challenge to lattice simulations where the cost of large grids is great.
 \bigskip \bigskip

\noindent{\bf Acknowledgements:} The authors are grateful for discussions with Kimmo Tuominen. NE is supported by an STFC Consolidated Grant and thanks ESF Holograv for additional funding. MS is supported by an STFC studentship.

\end{document}